\documentclass[12pt,notoc]{JHEP3}
\usepackage{amsmath,amssymb,euscript,array}
\newcommand{\startappendix}{
\setcounter{section}{0}
\renewcommand{\thesection}{\Alph{section}}}
\newcommand{\Appendix}[1]{
\refstepcounter{section}
\begin{flushleft}
{\large\bf Appendix \thesection: #1}
\end{flushleft}}
\usepackage{epsfig}

\def\Tr{{\rm Tr}}

\def\det{{\rm det}}

\def\Dbarslash{\,\,{\raise.15ex\hbox{/}\mkern-12mu {\bar\D}}}
\def\Dslash{\,\,{\raise.15ex\hbox{/}\mkern-12mu \D}}
\def\delslash{\,\,{\raise.15ex\hbox{/}\mkern-9mu \partial}}
\def\delbarslash{\,\,{\raise.15ex\hbox{/}\mkern-9mu {\bar\partial}}}

\newcommand{\EQ}[1]{\begin{equation} #1 \end{equation}}

\newcommand{\SP}[1]{\begin{equation}\begin{split} #1
\end{split}\end{equation}}

\title{ AdS
Taub-NUT space and the $O(N)$ vector model on a squashed 3-sphere}
\author{\\Mark D.Yonge\\Department of Physics,\\ University of Wales Swansea,\\
Singleton Park, \\
Swansea, SA2 8PP, UK.\\\\
E-mail: {370058@swan.ac.uk}}
\preprint{swat/06/480          \\hep-th/yymmnnn}
 \abstract{In this note, motivated by the Klebanov-Polyakov conjecture
   we investigate the strongly coupled $O(N)$ 
vector model at large $N$ on a squashed three-sphere and its
holographic relation to bulk gravity on asymptotically locally
$AdS_4$ spaces. We present analytical results for the action of the
field theory as the squashing parameter $\alpha\rightarrow-1$,
when the boundary becomes effectively one dimensional. 
The dual bulk geometry is AdS-Taub-NUT space in the
corresponding limit. In this limit
we solve the theory exactly and show that the action of the strongly
coupled  boundary  
theory scales as $\ln(1+\alpha)/ (1+\alpha)^2$. This result is remarkably
close to the $-1/(1+\alpha)^2$ scaling of the Einstein
gravity action for AdS-Taub-NUT space.  These results explain the
numerical agreement presented in hep-th/0503238, and the soft
logarithmic departure is interpreted as a prediction for the
contribution due to higher spin fields in the bulk $AdS_4$ geometry.
}
\begin{document}
\section{Introduction and Summary}

The AdS/CFT correspondence states that string theories in
asymptotically AdS spacetimes with $d$ dimensions are dual to
certain conformal field theories in $d-1$ dimensions
\cite{KP7,KP8,KP2}. Testing these dualities is in general difficult
because the theories involved are very complicated and are only
tractable in different limiting regions of parameter space. However
in \cite{KP1} Klebanov and Polyakov suggested that a simpler duality
exists between the large $N$ limit of the singlet sector of the
critical $O(N)$ vector model in three dimensions and the minimal
bosonic higher spin gauge theory in four dimensional
 Anti de Sitter space. In \cite{KP5} an extension of this duality was proposed between the $O(N)$ model on a squashed three sphere and the
 higher
  spin gauge theory on AdS Taub-NUT and AdS Taub-Bolt geometries with a phase transition occurring between the two on the gravitational side.
The squashed three sphere is an $S^1$ bundle over $S^2$ with metric
\EQ{ds^{2}=\frac{a^{2 }}{ 4}
\Big(\sigma_1^{2}+\sigma_2^{2}+\frac{\sigma_{3}^{2}}{1+\alpha}\Big)\,.}
Where the $\sigma_i$ are defined by:
\EQ{\sigma_1+i\sigma_2=e^{-i\psi}(d\theta+i\sin{\theta}d\phi)} and
\EQ{\sigma_3=d\phi+\cos{\theta}d\phi} The squashing parameter
$\alpha$ lies in the range \EQ{-1\leq\alpha <\infty} with $\alpha=0$
corresponding to the round three sphere. In the large $\alpha$ limit
the squashed sphere approaches the direct product space $S^2 \times
S^1$, and the periodicity of the $S^1$ fibre can be thought of as an
inverse temperature. The limit $\alpha\to -1$ is the limit of
extreme squashing which was not accessible analytically before, and
this will be the main focus of this work. In this limit one of the
dimensions becomes very large compared to the others and the field
theory becomes effectively one dimensional.

 This duality has the advantage compared to the usual string/gauge
 theory dualities in 
that the QFT is exactly solvable and can be compared to the
semiclassical properties of Einstein's gravity in the absence of a
proper formulation of the higher spin gauge theories in AdS Taub-NUT
and AdS Taub-Bolt spacetimes. It is useful to solve the $O(N)$ model
on a squashed three sphere because it provides a one parameter
family of field theory/ gravity dualities, whose free energies
exhibit a non monotonic behavior as a function of the squashing
parameter as argued in \cite{KP5}. For other related works on the
$O(N)$ model and the Klebanov-Polyakov duality, see
\cite{KP20,KP21,KP22}. 

The squashed three sphere is the conformal boundary of AdS Taub-NUT
and AdS Taub-Bolt geometries \cite{KP13,KP14}. As in the canonical
example of the Hawking Page transition\cite{KP6}, only one of these
two geometries dominates the partition function. In particular, as a
function of $\alpha$, there is a Hawking-Page transition from AdS
Taub-NUT to AdS Taub-Bolt, the latter dominating for large $\alpha$.
In \cite{KP10,KP11} the action of AdS Taub-NUT was found to
be:\EQ{I_{TN}=-\frac{6\pi}{GR}\frac{(1+2\alpha)}{(1+\alpha)^2}\label{ff1}}
where $G$ is Newton`s constant and $R$ is the Ricci scalar which is
negative in these backgrounds. For AdS Taub-Bolt the corresponding
result is:
\EQ{I_{TB}=\frac{24\pi}{RG}(1+\alpha)^{-\frac{1}{2}}
  (m_b+\frac{3}{4}r(1+\alpha)^{-1}-r^3)}    
with
\EQ{m_b=\frac{1}{2}r+\frac{1}{8r}(1+\alpha)^{-1}+\frac{1}{2}(r^3-\frac{3}{2}r(1+\alpha)^{-1}-\frac{3}{16r}(1+\alpha)^{-2})}
and
\EQ{r=\frac{1}{6}(1+\alpha)^{\frac{1}{2}}(1+(1-12(1+\alpha)^{-1}+9(1+\alpha)^{-2})^{\frac{1}{2}})}

In the limit of large $\alpha$ the AdS Taub-Bolt action grows
linearly:\EQ{I=\frac{4\pi}{9GR}\alpha\hspace{.5in}\alpha\rightarrow
\infty \label{linear}}
 The action of the $O(N)$ model was calculated  in \cite{KP5} for $\alpha>-\frac{8}{9}$ and
 is shown in Fig 1 below, and the result for the gravitational side is
 shown in Fig 2, where the action has been normalized so that it
 agrees with the field theory results at large $\alpha$ and a constant
 has been added so that the peaks coincide. They found a close
 numerical agreement 
 between the results for the $O(N)$ model and AdS Taub-NUT space below
 a critical value of $\alpha$ and with AdS Taub-Bolt above it, but
 with a smooth crossover between the two which suggests that the
 higher spin gauge fields have the effect of smoothing out the phase
 transition.

\begin{figure}[ht]
\centerline{\includegraphics[width=3in]{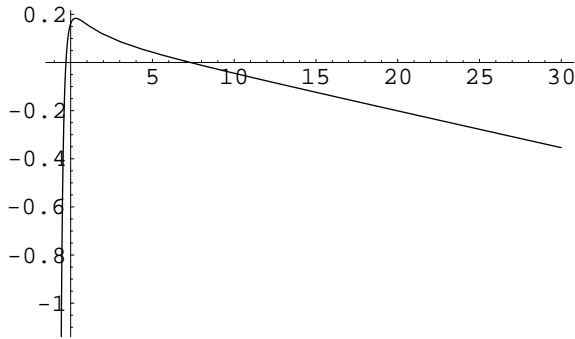}}
\caption{\footnotesize The action $\frac{I}{N}$ of the $O(N)$ model at strong coupling as a function of $\alpha$.} \label{actions}
\end{figure}

In \cite{KP5} the large $\alpha$ behavior of the QFT action  at
strong coupling was found to be:
\EQ{I_{a\lambda\gg1}=-\frac{N\zeta_{R}(3)}{10\pi^2}\alpha\label{large}}
The linear behaviour is guaranteed by the thermodynamic 
interpretation which emerges at large $\alpha$ wherein the
metric effectively approaches $S^2 \times S^1$.

\begin{figure}[ht]
\centerline{\includegraphics[width=3in]{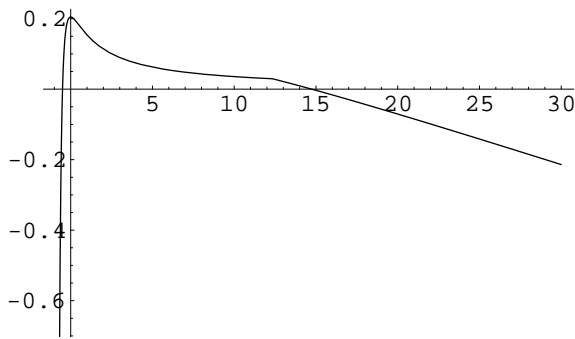}}
\caption{\footnotesize The Gravitational action as a function of
  $\alpha$.} \label{actions} 
\end{figure}

The striking and somewhat mysterious feature of the agreement found in
\cite{KP5}, is that the linear beahviour at large $\alpha$ turns 
over at small
$\alpha$ (see Figures 1 and 2). 
In this note we attempt to understand this non-monotonic
behaviour analytically, and in the process go beyond some aspectds of
the work of \cite{KP5} to the limit $\alpha\rightarrow
 -1$. The main result of this paper is the strong coupling action of
 the $O(N)$ model near the lower limit of the range of the squashing
 parameter, $\alpha \rightarrow -1$: 
 \EQ{I_{a\lambda\gg1}=\Big(\frac{\ln(1+\alpha)}{3(1+\alpha)^2}+\frac{0.0614093}{(1+\alpha)^2}\Big)N\label{ff2}}  
The results \eqref{large} and \eqref{ff2} are to be compared and
contrasted with  \eqref{linear} and \eqref{ff1}.

  It can be seen that the qualitative behavior of the free energy of the $O(N)$
 model as $\alpha\rightarrow-1$ and at large $\alpha$ closely
 reproduces the results of semiclassical gravity. The
 logarithmic deviation in the leading order term in the limit
 $\alpha\rightarrow-1$ is a prediction for the effect of including
 higher spin gauge fields in addition to gravity in the bulk
 dual. Interestingly 
 there appears to be no a priori reason why the results for the
 higher spin gauge theory should be so close to the pure gravity
 result, though these results suggest that the effects of the
 higher spin gauge fields cannot be drastic.

 In section two we summarize some useful results from the $O(N)$
 model and in section three we describe the calculation in more
 detail. Section four contains a discussion and summary.

 \section{O(N) Model in the Large N Limit}

The $O(N)$ model has been extensively studied in various dimensions e.g. see
\cite{KP15}. In Euclidean space the $O(N)$ model has the classical
action,

\EQ{S=\int dx^{D} \sqrt{g}\Big(\frac{1}{2}\nabla\Phi\cdot\nabla\Phi
+\frac{1}{2}m^{2}\Phi\cdot\Phi+\frac{\lambda}{4N}(\Phi\cdot\Phi)^{2}\Big)\,.}
The coupling constant $\lambda$ flows from a free fixed point in the UV to
another fixed point in the IR. This model can be solved exactly in
the strictly large $N$ limit by deriving an effective potential.
This can be done by introducing
 a homogenous background
expectation value $\phi$ for the $O(N)$ field and then splitting the
field into a VEV and fluctuations as follows:
 \EQ{\Phi=(\sqrt{N}\phi+\delta\phi,\pi_1,\pi_2,......,\pi_{N-1})\,.}
Here $\phi$ is the homogeneous background and $\delta\phi$ and
$\vec{\pi}$ are the fluctuations around it. Normally this  would
break the $O(N)$ symmetry to $O(N-1)$ resulting in goldstone
bosons, however as argued by \cite{KP5} in these circumstances the
symmetry is not broken because the path integral includes an
integration over the vacuum manifold which implies that symmetry
breaking does not occur in a compact space. The fluctuations can
then be integrated out, and the result is :
\SP{\frac{V_{\rm{eff}}(\phi,\sigma)}{N}=&\frac{1}{2}(m^{2}+\lambda\sigma)
  \phi^{2}-\frac{\lambda}{4}\sigma^{2}+\frac{1}{2{\rm Vol}(M)}
  \ln
\det^{\prime}\Big(\frac{-\Box+m^{2}+\lambda\sigma}{\mu^{2}}\Big)\\
& -\frac{1}{2{\rm
Vol}(M)}\Big(1+\ln{\pi}+\ln{\frac{\phi^{2}}{\mu}}\Big)\,,\label{v}} Where
Vol($M$) is the volume of the manifold on which the field theory is
formulated. In the present context $M$ represents the squashed three
sphere. $\mu$ is a dimensional scale which is like the sliding
renormalization scale. The prime in ${\rm det}^{\prime}$ indicates
that the integration was not done over the constant mode which is
then dealt with separately. In the large $N$ limit only the
configuration obtained by extremising \eqref{v} contributes to the
partition function. Minimizing the effective potential with respect
to $\phi$ and $\sigma$ yields the equations:
\EQ{\phi^2(m^{2}+\lambda\phi)=\frac{1}{{\rm Vol}(M)}\label{m1}} and
\EQ{\phi^2-\sigma+\frac{1}{{\rm Vol}(M)}{\rm Tr}^{\prime}\Big({\frac{1}
  {-\Box+m^2+\lambda\sigma}}\Big)= 0\label{m2}.}

An "effective pion mass" can then be defined:
\EQ{m^2_\pi=m^2+\lambda\sigma} so that equations \eqref{m1} and
\eqref{m2} can be rewritten as a gap equation for $m^2_\pi$
\EQ{m^2_\pi=m^2+\frac{\lambda}{{\rm Vol}(M)}{\rm Tr}\Big(\frac
{1}{-\Box+m^2_\pi}\Big)\label{gap}} where the constant mode has been
absorbed into the above. Once \eqref{gap} has been solved the
effective potential can be evaluated at the extremum to give the
action: \EQ{S=\frac{N}{2}\Big(-\frac{{\rm
Vol}(M)}{2\lambda}(m^2-m^2_{\pi^2})+\ln{
\det{\frac{-\Box+m^2_\pi}{\mu^2}}}+\ln({\mu^3 {\rm Vol} (M)})\Big).
\label{action}}
 To evaluate \eqref{gap} it is
necessary to evaluate the trace. This can be done by the method of
zeta function regularization. The zeta function for an elliptic
operator $A$ is defined by \EQ{\zeta(s)={\rm Tr}A^{-s}} so that
 \EQ{\ln{
\det{\frac{-\Box+m^2_\pi}{\mu^2}}}=-\lim_{s\to 0}{
\frac{d}{ds}\Tr(\frac{-\Box+m^2_\pi}{\mu^2})^{-s}} =
-\zeta^{\prime}(0).} The zeta function on the squashed three sphere
can be written in the form\cite{KP12,KP260,KP27} \EQ{\zeta
(s)=\sum_{l=1}^{\infty}\sum_{q=0}^{l-1}\frac{l(a\mu)^{2s}}{(l^2+\alpha(l-1-2q)^2+a^2m^2_\pi-1)^s.}\,,\label{zeta}}

The self-consistent gap equation which determines the solution of the
model is highly non-trivial for two reasons. Firstly, it is a
a nonlinear equation for $m_\pi^2$. Secondly, it involves a zeta function
on the squashed sphere, namely $\zeta_{m_\pi^2}(1)$ which is a
complicated object and needs to be defined via analytic
continuation. 

In addition to these ingredients, we need to specify 
the coupling constant of the
theory $\lambda$ which is dimensionful. The relevant dimensionless
parameter in the theory is the combination $a \lambda $. Since
$\lambda$ is a relevant coupling in three dimensions, $a\lambda\ll 1$
is the weak coupling limit corresponding to taking the sphere size to
be small, thus approaching the UV free fixed point.

We will be primarily interested in the strong coupling limit 
$a \lambda\rightarrow\infty$ which corresponds to the IR
fixed point theory on the squashed sphere. 
In this limit, a drastic simplification of the gap
equation occurs, allowing us to solve the problem analytically in
the $\alpha \rightarrow -1$ limit. 
The gap equation at strong coupling \eqref{gap} determines $m_\pi^2$
to be a zero of $\zeta_{m_\pi^2}(1)$. The results of \cite{KP5}
provide evidence that the resulting value for $m_\pi^2$ is finite and
non-negative for all allowed values of the squashing parameter $\alpha$.

We will now evaluate the action of the theory in the
$\alpha\rightarrow -1$ limit, at strong coupling $a\lambda\rightarrow \infty$.
The fact that $m_\pi^2$ has a finite value determined by the zero of
$\zeta_{m_\pi^2}(1)$, implies that the first term in \eqref{action} is
zero 
at strong coupling. The volume term will finally be found to give a
subleading contribution to the action. The dominant contribution in
the $\alpha \rightarrow -1$ limit therefore is:
$-\frac{N}{2}\zeta^{\prime}(0)$. As this is superficially divergent
it needs to be analytically continued by standard methods described
in the appendix.

\section{The $O(N)$ model as $\alpha\rightarrow -1$
}

The zeta function \eqref{zeta} is superficially divergent, but a
finite value may be obtained by analytically continuing the sum,
firstly by applying the Abel-Plana formula, and then by carrying out
a Sommerfield Watson transformation. The lengthy technical details are
summarized in the Appendix. 
We find:
\EQ{\frac{\zeta(s)}{\mu^{2s}}=\frac{a^{2s}A}{(1+\alpha)^s}+a^{2s}\int_{0}^{1}\frac{B(y)dy}{(1+\alpha
y^2)^s}-\frac{2ia^{2s}}{(1+\alpha)^s}\int_{0}^{\infty}\frac{C(y)dy}{1+\exp
(2\pi y)}} where the functions $A,B,C$ are themselves infinite sums 
defined in the Appendix.
After lengthy complex analysis manipulations, using the above
result, we find that the zeta function in the limit
$\alpha\rightarrow-1$, is given by:
\SP{\zeta^{\prime}(0)=&-\frac{\log(1+\alpha)}
  {3(1+\alpha)^2}-\frac{10}{9(1+\alpha)^2}-\frac{2\ln2}{3(1+\alpha)^2}\\
&+\int_{0}^{\infty}\frac{16y\ln(1+4y^2)-8(4y^2-1)\tan^{-1}{2y}}{\exp(2\pi
y)-1}dy .}

This then evaluates to:

\EQ{I=\frac{N}{2}\Big(\frac{\ln(1+\alpha)}{3(1+\alpha)^2}+\frac{0.0614093}{(1+\alpha)^2}\Big).}

The above argument relies on $a^2m_\pi^2$ being finite at strong
coupling and in the entire range of allowed values of $\alpha$. 
In fact the numerical
results in \cite{KP5} indicate that $a^2 m_\pi^2$ from the solution to
the 
strong coupling gap equation 
approaches zero in the $\alpha\rightarrow -1$ limit.

\section{Summary and Discussion}

We have solved the strongly coupled $O(N)$ model exactly in the
limit $\alpha\rightarrow-1$ and found a soft logarithmic deviation
from the results of semiclassical gravity in this regime. It is
surprising that the scaling of the action of the strongly
coupled $O(N)$ model in this limit, is so
similar to that of classical gravity on AdS-Taub-NUT space. 
The ``anomalous'' logarithmic deviation can only be explained
within the
confines of the Klebanov-Polyakov conjecture as being due to the
effects of the higher spin gauge fields. There appears to be no
obvious physical explanation for the behavior of the action in this
limit, but it would be interesting to see if it is because in the
$\alpha\rightarrow-1$ limit the field theory becomes effectively
one dimensional.

In the other extreme of the allowed range of $\alpha$, namely at large
$\alpha$, it is evident from \eqref{linear} and \eqref{large} that
classical gravity and boundary field theory are qualitatively
similar. This is not very surprising, since at large $\alpha$,
the boundary theory can be reinterpreted as being at a finite temperature
given by $\alpha$. The linear scaling of the action with $\alpha$, and
equivalently the free energy scaling as $\alpha^2$, is what one
expects in a field theory in three dimensions. Nevertheless, from this
we learn, assuming the validity of the Klebanov-Polyakov conjecture,
that the higher spin theory dual to the $O(N)$ model at large
squashing, should behave
in qualitatively the same fashion as Einstein gravity in AdS-Taub-Bolt
space. We remark that the coefficients for the field theory \eqref{large}
and gravity \eqref{linear} actions are not expected to match as the
higher spin gauge 
fields were not included in the gravity calculation. In any case,
matching of these coefficients only sets up the dictionary between
$1/N$ in field theory and the bulk curvature in units of the 4d Planck mass.
The above discussion 
compares with the $AdS_5/CFT_4$ case where doing a strongly coupled
field theory calculation is difficult and there is a $3/4$
discrepancy factor between strong and weak 't Hooft 
coupling results due to higher stringy modes becoming light at 
large, string-scale curvatures in the string dual of the weakly coupled
gauge theory.

The analytic results obtained in this paper,
for the strongly coupled field theory near $\alpha\rightarrow -1$,
when combined with the linear behaviour at large $\alpha$, reproduce remarkably
well the non-monotonic behaviour of the classical bulk gravity action
presented in Figure 2. Note that the non-monotonic behaviour in the
bulk (without higher spin fields) is due to a
Hawking-Page transition which is necessary in order to pass over from
the  AdS-Taub-NUT to the AdS-Taub-Bolt phase, the latter showing a
linear beahviour with $\alpha$ at large $\alpha$. The  
message is that even though we don't have a proper formulation of
the higher spin theory in these backgrounds, our results suggest
that gravity reproduces qualitatively similar results to the higher
spin gauge theory dual to the $O(N)$ model.

Finally, AdS Taub Nut space is obtained by filling the volume of a squashed
three sphere with a hyperbolic metric with negative cosmological
constant\cite{KP16}. In the limit $\alpha\rightarrow-1$ the space
becomes a Bergmann space which can be described as a coset space
$SU(2,1)/U(2)$ which has been studied in \cite{KP17}. It would be
interesting to understand the behavior of the action from the bulk
perspective by considering higher spin gauge fields on this Bergmann
space. For a detailed construction of bulk-boundary and bulk-bulk
propagators in this space see \cite{KP23}. Other work in AdS Taub
Nut space is contained in \cite{KP24,KP25,KP26}.

\textbf{Acknowledgements}

We would like to thank Prem Kumar and Carlos Nunez for several
useful discussions during this work. Mark Yonge would like to thank
PPARC for their financial support during this work.

\startappendix \Appendix{Analytic continuation of the zeta function}

As \eqref{zeta} is divergent if $s$ is set directly to zero it will
need to be continued analytically. This was done by converting the
sum over $q$ into an integral using the Abel-Plana formula and
evaluating the $l$ summation using a Sommerfield-Watson
transformation. \eqref{zeta} has branch cuts at
\EQ{q=\frac{l-1}{2}\pm\frac{1}{2}
  \Big(\frac{1-l^2-a^2m^2_\pi}{\alpha}\Big)^{1/2}\,,\label{cut}}
with these branch cuts the Abel-Plana formula of the form
\eqref{plana} may be used to evaluate the $l$ sum.
\SP{\sum_{i=n}^{m}\phi(x)=&\frac{1}{2}(\phi(n)+\phi(m))+
\int_{n}^{m}\phi(x)dx\\& -i\int_{0}^{\infty}\frac{dy}{\exp(2\pi y)-1}
(\phi(n-iy)-\phi(n+iy)-\phi(m-iy)+\phi(m+iy))\,.{\label{plana}}}

Using this we obtain:
\EQ{\frac{\zeta(s)}{\mu^{2s}}=\frac{a^{2s}A}{(1+\alpha)^s}+a^{2s}\int_{0}^{1}\frac{B(y)dy}{(1+\alpha
y^2)^s}-\frac{2ia^{2s}}{(1+\alpha)^s}\int_{0}^{\infty}\frac{C(y)dy}{\exp
(2\pi y)-1}\,,{\label{base}}} where:
\EQ{A=\sum_{l=1}^{\infty}\frac{l}{((l+G)^2-h^2)^s}\,,}
\EQ{B=\sum_{l=1}^{\infty}\frac{l(l-1)}{((l+I)^2-J^2)^s}\,,}

\EQ{C=\sum_{l=1}^{\infty}l(\frac{1}{((l+K)^2-M^2)^s}-\frac{1}{((l+K^*)^2-(M^2)^*)^s}\,.}
Here $G,H,I,J,K,M$ are given by:

\EQ{G=\frac{-\alpha}{1+\alpha}}
\EQ{-H^2=\frac{a^2(m_\pi)^2(1+\alpha)-1}{(1+\alpha y^2)^2}\,,}
\EQ{I=\frac{-\alpha y^2}{(1+\alpha y^2)}\,,}
\EQ{-J^2=\frac{a^2m_{\pi}^2(1+\alpha y^2)-1}{(1+\alpha y^2)^2}\,,}
\EQ{K=\frac{\alpha (-1+2iy)}{1+\alpha}\,,} \EQ{-M^2=\frac{a^2
m_{\pi}^2 -1 +\alpha (-1+2iy)}{1+\alpha}-\frac{\alpha^2
(-1+2iy)^2}{(1+\alpha)^2}\,,}

The sums over $l$ can be evaluated using a Sommerfield-Watson
transformation. Apply this to the $A$ we find
\EQ{A=\frac{i}{2}\int_{C_1}\frac{z\cot{\pi z}
dz}{((z+G)^2-H^2)^s}\,,} where the contour $C_1$ is shown in figure
three. For ${\rm Re} (s)>2$ this can be deformed into $C_2$, also shown
in figure three.
\begin{figure}[ht]
\centerline{\includegraphics[width=3in]{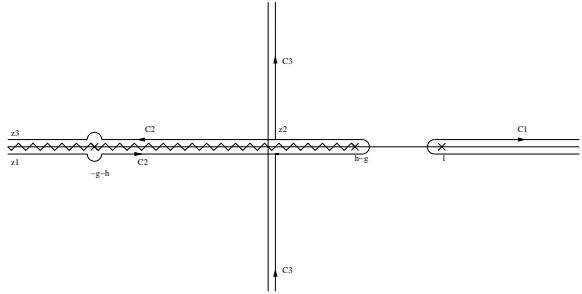}}
\caption{\footnotesize The contours used for the analytic
continuation of the zeta function.}untitled folder \label{actions}
\end{figure}

It is useful to rewrite $\cot{\pi z}$ with the identities
\EQ{\cot{\pi z}=i\Big(1+\frac{2}{\exp{2i\pi z-1}}\Big)\,,} between
$z_1$ and $z_2$ and \EQ{\cot{\pi z}=i\Big(-1+\frac{2}{1-\exp(2i\pi
z)}\Big)\,,} between $z_2$ and$z_3$.

The integrals over the exponential pieces are then manifestly finite
and can be evaluated along $C_3$. The integrals over the constant
pieces can be done analytically for ${\rm Re} (s)>2$ . These
expressions are then evaluated at $s=0$ to define the analytically
continued function. A similar method is used for $B$ and $C$. In the
case of $C$ the branch points are not on the real axis so the
integrals are no longer along the real axis but along a tilted
contour. Using this method the following results are obtained in the
limit $\alpha$ tends to minus one.:
\EQ{A\mid_{s=0}=\frac{1}{(1+\alpha)^2}\,,}
\EQ{\frac{d}{ds}A\mid_{s=0}=\frac{2\log(1+\alpha)}
  {(1+\alpha)^2}+\frac{3}{(1+\alpha)^2}-\frac{2\log{2}}{(1+\alpha)^2}\,,}

\EQ{\int_{0}^{1}B\mid_{s=0}(2\log{a}-\log(1+\alpha
y^2)=-\frac{4\log{a}}{3(1+\alpha)^2}+\frac{1}{3(1+\alpha)^2}+\frac{2
\log(1+\alpha)}{3(1+\alpha)^2}\,,}

\EQ{\int_{0}^{1}\frac{dB}{ds}\mid_{s=0}=\frac{4\log(1+\alpha)}
  {3(1+\alpha)^2}-\frac{10}{9(1+\alpha)^2}-\frac{4\log{2}}{3(1+\alpha)^2}}
\EQ{C\mid_{s=0}=-\frac{8yi}{(1+\alpha)^2}\,,}

\EQ{\frac{dC}{ds}\mid_{s=0}=\frac{2i}{(1+\alpha)^2}
  (4y(-7+8\log{2}-4\log(1+\alpha)+2\log(1+4y^2))-4(4y^2-1)\tan^{-1}{2y})\,.}
Putting the above together gives:
\SP{\zeta^{\prime}(0)=&-\frac{\log(1+\alpha)}
  {3(1+\alpha)^2}-\frac{10}{9(1+\alpha)^2}-\frac{2\ln2}{3(1+\alpha)^2}\\
&+\int_{0}^{\infty}\frac{16y\ln(1+4y^2)-8(4y^2-1)\tan^{-1}{2y}}{\exp(2\pi
y)-1}dy .}

The integrals can then be evaluated numerically.

\end{document}